\documentclass[english]{article}
\usepackage[T1]{fontenc}
\usepackage[latin9]{inputenc}
\usepackage{graphicx}
\usepackage{esint}

\makeatletter
\newcommand{\lyxaddress}[1]{
	\par {\raggedright #1
	\vspace{1.4em}
	\noindent\par}
}

\usepackage{babel}

\makeatother

\usepackage{babel}
\begin{document}
\title{Universal features in lifetime distribution of clusters in hydrogen
bonding liquids}
\author{Ivo Juki\'{c}$^{\dagger}{}^{,\ddagger}$ , Bernarda Lovrin\v{c}evi\'{c}$^{\ddagger}$\thanks{bernarda@pmfst.hr},
Martina Požar$^{\ddagger}$, and Aurélien Perera$^{\dagger}$\thanks{corresponding author (aup@lptmc.jussieu.fr)}}
\maketitle

\lyxaddress{$^{\dagger}$Laboratoire de Physique Théorique de la Matière Condensée
(UMR CNRS 7600), Sorbonne Université, 4 Place Jussieu, F75252, Paris
cedex 05, France.}

\lyxaddress{$^{\ddagger}$Department of Physics, Faculty of Sciences, University
of Split, Ru\dj era Boškovi\'{c}a 37, 21000, Split, Croatia.}
\begin{abstract}

Hydrogen bonding liquids, typically water and alcohols, are known to form labile structures (network, chains, etc...), hence the lifetime of such structures is an important microscopic parameter, which can be calculated in computer simulations. Since these cluster entities are mostly statistical in nature, one would expect that, in the short time regime, their lifetime distribution would be a broad Gaussian-like function of time, with a single maximum representing their mean lifetime, and weakly dependent on criteria such as the bonding distance and angle, much similarly to non-hydrogen bonding simple liquids, while the long time part is known to have some power law dependence. Unexpectedly, all the hydrogen bonding liquids studied herein, namely water and alcohols, display highly hierarchic three types of specific lifetimes, in the sub-picosecond range 0-0.5ps The dominant lifetime very strongly depends on the bonding distance criterion and is related to hydrogen bonded pairs. This mode is absent in non-H-bonding simple liquids. The secondary and tertiary mean lifetimes are related to clusters, and are nearly independent on the bonding criterion. Of these two lifetimes, only the first one can be related to that of simple liquids, which poses the question of the nature of the third life time.
The study of acohols reveals that this 3rd lifetime is related to the topology of H-bonded clusters, and that its distribution may be also affected by the alkyl tail surrouding ?bath?. This study reveals that hydrogen bonding liquids have a universal hierarchy of hydrogen bonding lifetimes with a timescale regularity across very different types, and which depend on the topology of the cluster structures
\end{abstract}

\section{Introduction}

Labile structures in associated liquids and mixtures pose the problem
of the role of kinetics of said structures, and its influence on the
thermophysical and dynamical properties of these systems 
\cite{Labile_complex_kinetics,Paul2017}.
Such structures play an important role in soft matter, as for example
with micelles and lamellae \cite{Micelle_Glatter,De1998,Lamellae},
and more particularly in biology, wherein the labile character of
various functional molecular entities, such as enzymes for example,
wears a fundamental operational nature \cite{Labile_enzyme_1,Labile_enzyme_2,Labile_enzyme_metal}.
One might even postulate that biological systems have been built by
the increasing role played by such labile structures in the early
evolution of primitive biochemical systems
\cite{Trevors2001,Deamer2006}. Based on these premises,
it is important to better understand the interplay between labile
nature of molecular assemblies and the role of kinetics in their dynamics
\cite{Labile_complex_kinetics,Labile_enzyme_1,Labile_enzyme_2,Labile_enzyme_metal}.

A first step in that direction would be analyzing
the life time distribution of the hydrogen bonding process which is
at the root of molecular association \cite{2019_Propylamine2,2020_Ethanol_PCCP}.
Since H-bonding is essentially a quantum mechanical process, it is
affected by various intra molecular motions, and in turn it affects
inter molecular motions. It is not clear if the various experimental
techniques, which allow to probe the frequencies associated with these
motions, can unambiguously answer the question posed above \cite{Spectro_IR_HB_alc,Spectro_Murthy,Spectro_IR_Craven_Amines,Spectro_IR_Craven_Alcohols}.
Since this is a many body quantum mechanical phenomenon, it is not
even clear if approximate theories can provide an alternative approach
to this question. On the other hand, computer simulation provide a
direct access to the statistics of molecular motions, and are able
to answer this question. One could even answer this question at the
level of classical physics, where the hydrogen bond is modeled by
the Coulomb pairing of opposite charges \cite{TIP_Methanol_HB_analysis,Starr2000},
and whose pertinence has been amply proven by more than fifty years
of computer simulations and force field development.

In fact, the question posed above has been already partly answered
by Luzar and Chandler in their 1996 Nature paper \cite{Luzar1996}.
In this paper, the authors focus essentially in the long time
kinetics of H-bonding in water, beyond the initial 0.5 ps which they
mention as the transient regime. This choice is amenable to a theoretical
approach of the H-bonding kinetics, which is shown to be non-exponential,
and further supported by classical computer simulations, but it does
not explain the origin of the transient behaviour observed at short
times. The present work aims at revealing surprising repeatability
of this transient part across several H-bonding liquids. It is initially
motivated by the fact that H-bonding liquids other than water, such
as alcohols and amines, are known to form short chain-like clusters,
both from scattering experiments \cite{ExpScattOldAlc,ExpScattOldAlc2,
ExpScattMaginiMethanol,
ExpScattNartenEthMeth,ExpScattFinnsMonools,ExpScattJoarderAlcohols,
EXP_SANS_Soper_pure_methanol,ExpScattBenmoreEthanol,ExpScattMatijaMonools,
EXP_Sarkar_1propanol,EXP_Pustai_MethEthProp,Tomsic_butanol,2019_Propylamine2,2020_Neat_Alc_Germans},
spectroscopy investigations \cite{EXP_Spectro_Methanol_clusters_2,
EXP_Spectro_Ethanol_clusters_2,
EXP_Spectro_alcohol_mixtures_clusters,EXP_Spectro_Methanol_clusters,
EXP_Spectro_ethanol_clusters}
and computer simulations \cite{SIM_Bako_methanol_EXP,SIM_LudwigMethanol,AUP_Neat_Alcohols_JCP,
AUP_Neat_alcohols_PRE,SIM_Finci2_linear_alcohols,2016_JCP_ethMeth,2017_Propylamine1}.
Exploring this transient regime, essentially in the sub 0.5 ps region,
we have uncovered a universal dynamical behaviour common to all H-bonding
liquids, which is unexpected in this time domain, where the differences
in the molecular interactions play an important role. Indeed, such
universality would be more expected in the long time kinetics regime,
where common features of the association process are likely to settle
\cite{Paul2017,Schreiber2009}.

In order to better appreciate the results presented herein, it is
useful to picture the H-bonding process as essentially a random process
at short times and distances, governed by the random molecular encounters.
Intuitively, one would expect a broad distribution of H-bonding life
times, as function of both time and H-bonding distance, centered around
some mean representative life time, which could be subsequently searched
in the various experimental techniques investigating the relaxation
processes. What we uncover here, is that while the distribution of
H-bonding distances are indeed as hypothesized, the life time distributions
exhibit several specific times, which vary from $1$ at very short
distances to $3$ at larger ones, where the distances are picked around
the main peak of the oxygen-oxygen distribution function, typically
in the range $2.5\mathring{A}$ to $3.5\mathring{A}$, this latter
value being often used in the literature \cite{SIM_Ferraio_Methanol,SIM_Guardia_Wat_Alc,SIM_Skarmoutsos_Guardia_ethanol,
SIM_EXP_TBAWat_Matija},
as in the Luzar paper mentioned above \cite{Luzar1996}. These specific
times, and related distribution, appear as common to all three associated
liquids we have investigated herein, namely water, alcohols and amine,
and across different force field representation. In contrast, the
long time kinetics show specificities related to each type of liquid.
Although we have used classical force field, this universality strongly
suggests that these times should be a real feature of the associated
liquids. We suggest that these specific times correspond to three
types of molecular associations: dimer, linear chain-like clusters
and other types of clusters. This finding suggests that self-assembled
labile structures have typical life times related to their differences
in microscopic topology. In that, it helps understand how such structures,
when complexified by appropriate molecular entities, could acquire
an important role in the pre-biotic phenomena. More importantly, it
shifts the interest to such structures, from the usual long time kinetics
approach, where kinetic constants play an important role, to short
times and distance, in the range of which new self assembled objects
appear, and could possibly play the role of new molecular species
in a very restricted spatio-temporal region.

\section{Theoretical, models and simulation details}

We would like to first stress that the present manuscript deals with
the H-bond lifetime distribution itself, and not the time auto-correlation
of it, which has been the subject of previous works by other authors
\cite{Luzar1996,Luzar2000,Galamba,Voloshin2009,GeigerStanley}.
It is therefore important to clarify the differences in the theoretical
backgrounds. The hydrogen bond is a real property of many associated
liquids in many contexts. However, for calculational purposes, it
is necessary do define the two following parameters, the bonding distance
$r_{c}$ between the two donor/acceptor atoms, and the corresponding
bonding angle $\theta_{c}$. In computer simulations, two molecules
$i$ and $j$ are considered as H-bonded, when the distance $r_{ij}$
between the corresponding donor/acceptor atoms $A_{i}$ and $B_{j}$
obeys $r_{ij}\leq r_{c}$, and the angle $\theta_{ij}=\widehat{A_{i}HB_{j}}$
obeys $|\theta_{ij}|<\theta_{c}$. The $A_{i}$ and $B_{j}$ atoms
are typically oxygen atoms, such as in water, but they can also refer
to nitrogen atoms, such as for 1-propylamine considered in this work.
Below, we will refer as $\mathcal{C}$ the ensemble of atoms which
verify both criteria. 
For each pair of molecules in this time interval, there are 2 characteristic
times: that $t_{ij}$ when they first bond and $\tau_{ij}$ when they
break apart for the first time. For this pair, we define a time dependant
random variable $H_{ij}(t)$, such that 
\begin{equation}
H_{ij}(t)=H(t-t_{ij})H(\tau_{ij}-t)\label{Hij}
\end{equation}
where $H(t$) is the Heaviside function, and $H_{ij}(t)=1$ for $t_{ij}<t<\tau_{ij}$
and zero elsewhere.

We believe that this is the first proper definition
of the function $h(t)$ introduced in the past literature 
\cite{Luzar1996,Luzar2000,Galamba,Voloshin2009,GeigerStanley}.
 From this random variable one can measure
several statistical averages and correlation, and in particular the
auto-correlation function $c(t)=<h(0)h(t)>$ which has been studied
in the past.

In the present work, we focus on the lifetime distribution itself,
and the appropriate random variable is $h_{ij}(t)$ can defined from
$H_{ij}(t)$. 
We first take the derivative $dH_{ij}(t)/dt=\delta(t-t_{ij})-\delta(t-\tau_{ij})$,
and remove the origin part. Then we define $h_{ij}(t)$ as gauge variable
(because of the derivative/integration operations) related to $H_{ij}(t)$
through

\begin{equation}
h_{ij}(t)=\int dt\delta(t-\tau_{ij})\label{hij}
\end{equation}
which is $1$ when the Hbond breaks at time $\tau_{ij}$ and zero
elsewhere.

This variable is therefore adapted to built the lifetime histogram,
and the associated lifetime distribution defined as

\begin{equation}
L(t)=\frac{1}{T_{0}L_{0}}\sum_{ij\in\mathcal{C}}h_{ij}(t)\label{Lt}
\end{equation}
where the normalisation factor $L_{0}$ is defined as

\begin{equation}
L_{0}=\frac{1}{T_{0}}\int_{0}^{T_{0}}dt\left[\sum_{ij\in\mathcal{C}}h_{ij}(t)\right]\label{NLt}
\end{equation}
 It is easily verified that $L(t)$ is a probability distribution
which verifies

\begin{equation}
\int_{0}^{T_{0}}dtL(t)=1\label{PLt}
\end{equation}
Eqs.(\ref{Lt},\ref{NLt}) provide a direct computational indication
as how to evaluate $L(t)$ in a given computer simulation. An average
over all possible time origin is implict. The auxiliary Gromacs program
\emph{gmx hbon}d module with the \emph{-life} option, allows to compute
$L(t)$. We have also checked through our own code that it was consistent
with the definitions given above. It should be noted that one can
end the H-bonding as soon as two bonded atoms part away according to
the chosen criterion. This is the strict definition, which has been
adopted in this work, but also in Gromacs. However, in reality, a
broken bond could be reformed quickly, and perhaps some margin should
be allowed for rebinding, which could be added to the ensemble of
criterion in $\mathcal{C}$. This could be conveniently introduced
by replacing the Dirac delta in Eq.(\ref{hij}) by a Gaussian function,
which we will consider in another context.

This study focuses on the typical hydrogen bonding liquids, which
are water and alcohols, which are both based on the OH group. Mono-ols
such as methanol, ethanol and propanol have been studied. The SPC/E
\cite{FF_SPCe} and TIP4P\_2005 \cite{FF_TIP4P_2005} models were
used to simulate water. Alcohols were modeled with the OPLS-UA \cite{FF_OPLS_alcohols_1}
and TRaPPE-UA \cite{FF_Trappe_Alcohols} forcefields. 

The program package Gromacs, version 2018.1 \cite{MD_gromacs_4_5,MD_gromacs_parallelization}
was used for all molecular dynamics simulations. The simulation protocol
has been the same for all neat systems. The initial configurations
of N molecules were generated by random molecular positioning with
the program Packmol \cite{MD_Packmol}, which were then energy minimized.
The values for N considered here are typically N=1000 or N=2048. The
systems were then equilibrated in the isobaric constant NpT ensemble
for at least 5 ns, followed by a production run of at least 5 ns.
Finally, an additional production run of 300 ps was performed to gather
sufficient data for the analysis of dynamic quantities (every configuration
was sampled).

The integration algorithm of choice was the leap-frog \cite{MD_INT_Leapfrog}
and the time step was 2 fs. The electrostatics were handled with the
PME method \cite{MD_PME} and the constraints with the LINCS algorithm
\cite{MD_constraint_LINCS}. The short-range interactions were calculated
within the 1.5 nm cut-off radius. Each neat liquid was simulated at
ambient conditions. Temperature was maintained at T = 300 K using
the Nose--Hoover \cite{MD_thermo_Nose,MD_thermo_Hoover} thermostat,
while pressure was kept at p = 1 bar with the Parrinello--Rahman
barostat \cite{MD_barostat_Parrinello_Rahman_1,MD_barostat_Parrinello_Rahman_2}.
The temperature algorithms had a time constant of 0.2 ps, while the
pressure algorithm was set at 2 ps.

The lifetime calculation is part of the post simulation analysis Gromacs
package (\textit{gmx H-bond} module) . It follows the usual process
of monitoring the lifetime of pairs of H-bonded O atoms across different
water molecules. The O-O H-bonding distance $r_{c}$ and the O-H-O
angle are input parameters. In this study, while we vary the bonding
distance, the angle is maintained to the accepted value of $180^{\circ}\pm30$.
We have verified that there is much less dependence on the angular
bias than the distance $r_{c}$ .

\section{Results }

Usually, H-bonding is discussed in terms of the spatial distribution,
through pair correlation functions and their spatial representations,
as proven by their overwhelming representation in the literature,
as compared with their temporal one. Therefore, the finding of a corresponding
temporal universality can be considered as an important feature, which
should help better understand the role of labile structures.

\subsection{Static structural properties of H-bonding}

The universality in kinetics claimed in this work is supported by
the static properties, in particular by the structural properties.
Fig.1 shows the various oxygen-oxygen pair correlation functions $g_{OO}(r)$
obtained from computer simulations of the SPC/E water model, and first
two OPLS alcohol models. The first peaks, shown in the main panel,
demonstrate that the contact pairing is dominated by the oxygen atom
size as well as the strong H-bonding pairing induced by the O-H-O Coulomb
association. It is important to note that these first peaks are relatively
robust across models, such that the distributions of bonding distances
do not vary much across models. This is further enforced by the striking
resemblance of distribution times across models, as shown below and
in the SI material.

\begin{figure}[ht]
\centering \includegraphics[height=7cm]{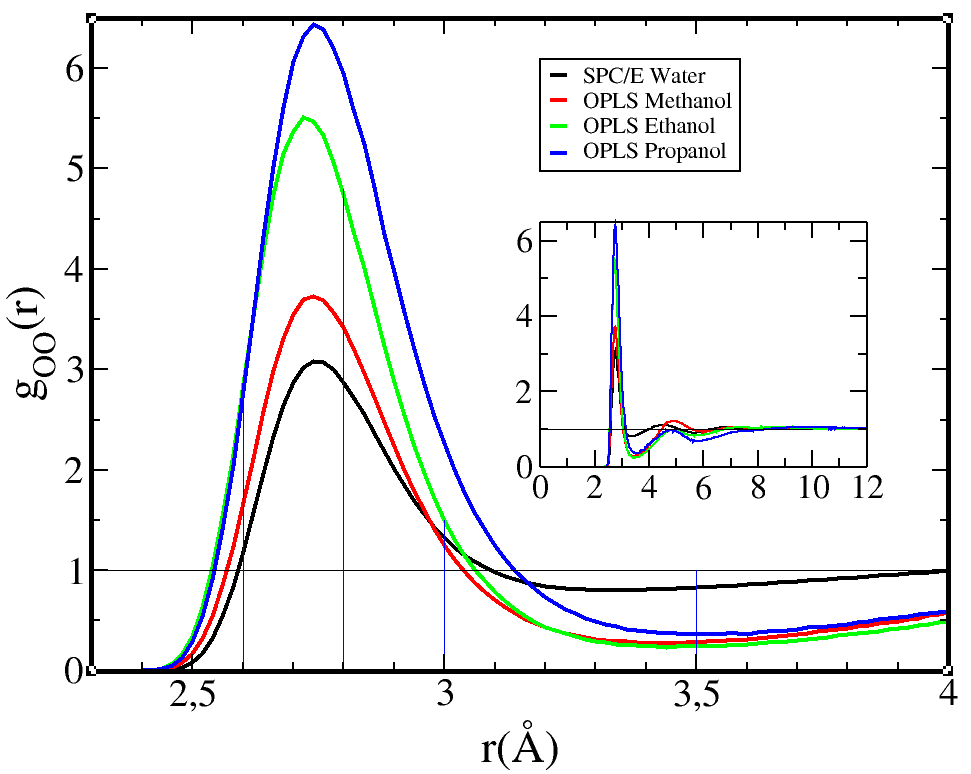} 
\caption{Oxygen-oxygen distribution functions $g_{OO}(r)$ for water and first alcohols.
The vertical blue lines represent a sample of the H-bonding distances
used in this work. The inset represent a wider range of these functions. }
\label{fig1-gr} 
\end{figure}

We turn to the H-bond donor-acceptor distance distributions, which
is shown in Fig.2, for the same models as in Fig.1. We immediately
note the expected feature discussed in the Introduction, namely that
the distribution is indeed a broad one, centered around a distance
which corresponds to the central peak in Fig.1, around $2.7\mathring{A}$.
It confirms that the underlying bonding is essentially a random phenomenon
with no particular spatial specificity. This finding is somewhat in
contrast with what snapshots of the cluster analysis reveal, where
such clusters in water appear in various patterns, while for alcohols
it has been proven that chains and loops are predominant. 

\begin{figure}[ht]
\centering \includegraphics[height=7cm]{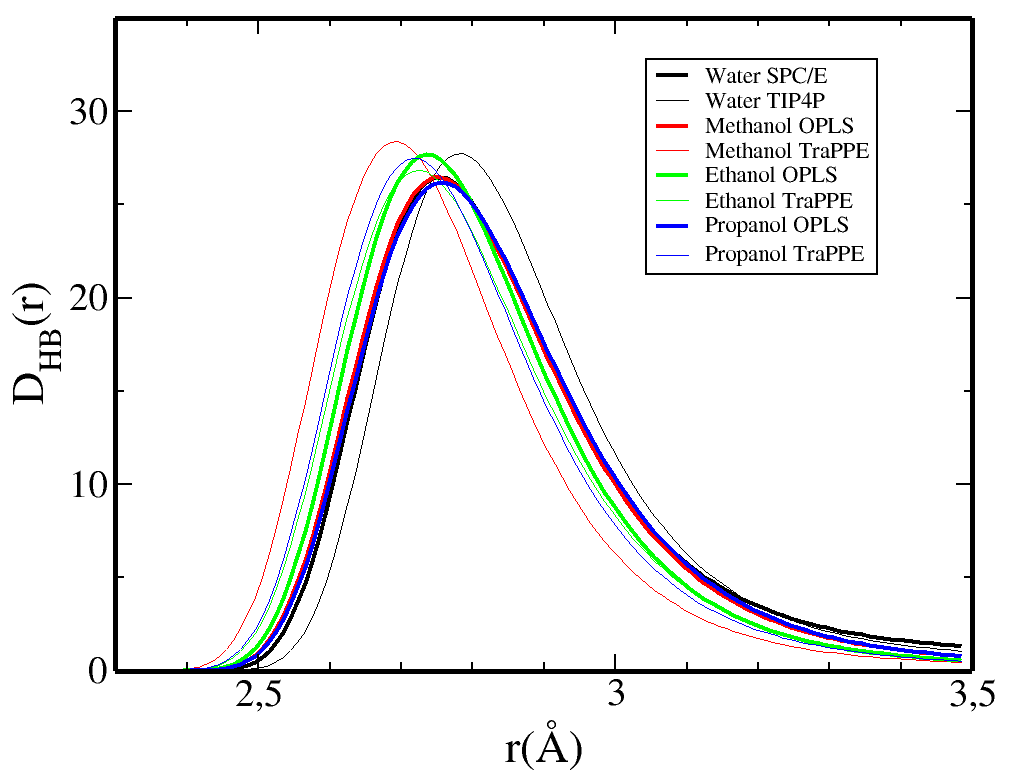} 
\caption{H-bond spatial distribution function $D_{HB}(r)$ for the different
models studied herein.}
\label{fig2-sHb}
\end{figure}

The curves in Fig.2 tend to indicate that donor-acceptor distribution
is essentially dominated by the dimers. Differences in clustering
are found in the larger distance in $g_{OO}(r)$, which is shown in
Fig.1. Indeed, it is seen in the inset of Fig.1 that the $g_{OO}(r)$
of the alcohols appears more depleted in the distance range $4-6\mathring{A}$,
which indicates that chain clusters are predominant in alcohols \cite{AUP_Neat_Alcohols_JCP,AUP_Neat_alcohols_PRE,2020_Neat_Alc_Germans}.
A direct consequence of this difference in clustering is the fact
that the scattering intensities of alcohols differ markedly 
from that of water, since they exhibit a pre-peak. \cite{ExpScattMaginiMethanol,ExpScattNartenEthMeth,ExpScattFinnsMonools,
ExpScattJoarderAlcohols,ExpScattMatijaMonools,2020_Neat_Alc_Germans}.
Perhaps the next most striking feature in Fig.2 is the near exact
superposition of all the curves, indicating that the H-bonding between
oxygen atoms is near invariant across molecular species. Small variations
are attributable to parameter differences in the underlying classical
force field, such as the partial charges and atom Lennard-Jones parameters.

The analysis above clearly indicates that the clustering differences
between water and alcohols are not seen in the spatial H-bond distribution,
which in a way is disappointing. But, we show now that this is not the
case for the temporal distributions.

\subsection{H-bond life times}

\subsubsection{Water}

Fig.3 shows the H-bond lifetimes distributions $L_{HB}(t)$ for the
SPC/E water model, in log-scale, for different distances encoded in
the color codes of the different curves, and ranging from $r_{c}=2.5\mathring{A}$
to $r_{c}=3.5\mathring{A}$ , which cover the distances from under
the first peak in Fig.1 until the first minimum. In terms of the corresponding
oxygen-oxygen potential of mean force, these correspond to the first
neighbour ranges. Fig.3 shows two remarkable features. First, for
small distance $r_{c}<3.0\mathring{A}$, one sees a series of peaks,
centered around a single maximum which represents the mean bonding
lifetime, and these peaks are seen to shift to larger times as $r_{c}$
is increased, as well as decreasing in magnitude. The shift to larger times
can be explained as the bonding distance criteria is brought closer
to the first maximum of $g(r)$ and beyond, one increases the probability
of pair association stability, hence the mean lifetime. 

\begin{figure}[ht]
\centering \includegraphics[height=7cm]{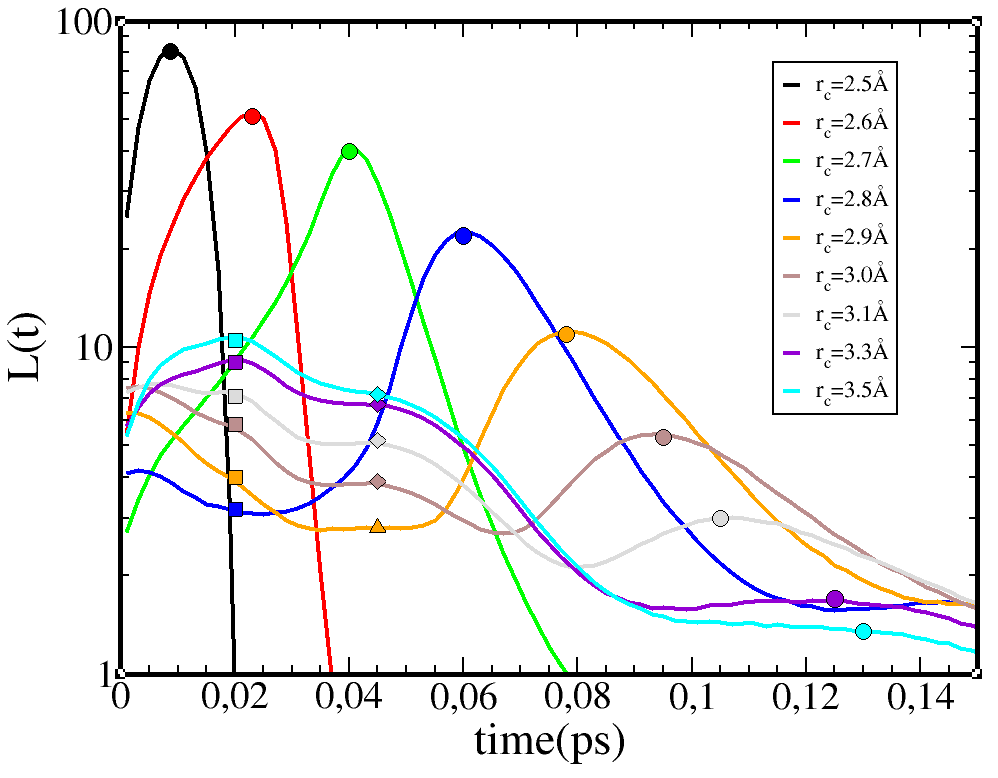} 
\caption{H-bond life time distribution $L(t)$ for SPC/E water, with different H-bonding
distances $r_{c}$. The symbols on each curve signal the corresponding peak position (see text): dot for first peak, square for second
and triangle for third. }
\label{fig3-LHb}
\end{figure}

However, as the contact distance is increased, we witness the appearance
of 2 secondary peaks, which grow at shorter times than the corresponding first peak lifetime. 
For example, for the mean H-bonding distance of $r_{c}=2.7\mathring{A}$
(green curve), the main peak is at 40 fs, while a broad shoulder at
15 fs witness the growth of the two secondary lifetimes. For the largest H-bonding
distance of $r_{c}=3.5\mathring{A}$ (in cyan), the secondary peaks are respectively
at 20fs and 45fs, and their amplitude is far superior to that of the
first peak/shoulder at 130fs, since they are respectively of 10 and
7, almost one order of magnitude larger that the for the first peak. In addition, we see that
these two secondary peaks are nearly always at the same respective
times of 20fs and 45fs. Anticipating the corresponding analysis for
the alcohols, we infer that these 2 secondary peaks must correspond
to lifetimes of bonded pairs within larger clusters, hence are signatures of
such clusters. The overall picture
which emerges from these findings is that, tightly H-bound molecules
(small $r_c$ ) have high probability but short life times,
but less tightly H-bound molecules (larger $r_c$) are less probable, but live
longer since they belong to a cluster. 
But a new problem
appears: if the secondary peaks corresponds to H-bonding within a cluster,
why two such peaks? Following the same inferring approach, we deduce 
that two types of clusters are present. From the analysis of the cluster
shapes in water, and also in alcohols, we infer that these peaks must correspond 
respectively to linear and  non-linear (globular, or branched) clusters.

Figs.SI-1 and SI-2 of the SI material show that the shape of these life times
is very similar across water models. It must therefore be a genuine
physical property of real water. 


Finally, we note that the conclusions inferred from the analysis above
are in contrast to a generic intuitive idea: that clusters could be
made of highly and tightly H-bound molecules. The picture which emerges
from the present analysis is that larger clusters are less probable than
dimers,
but they are also live longer. At the opposite range, dimer clusters
are highly likely, but they also break faster than larger clusters. Somewhere in between
these two extremes, one must have clusters which witness the existence
of labile entities, and the 3 peaks of the present finding correspond
to typical such clusters.

\begin{figure}[ht]
\centering \includegraphics[height=7cm]{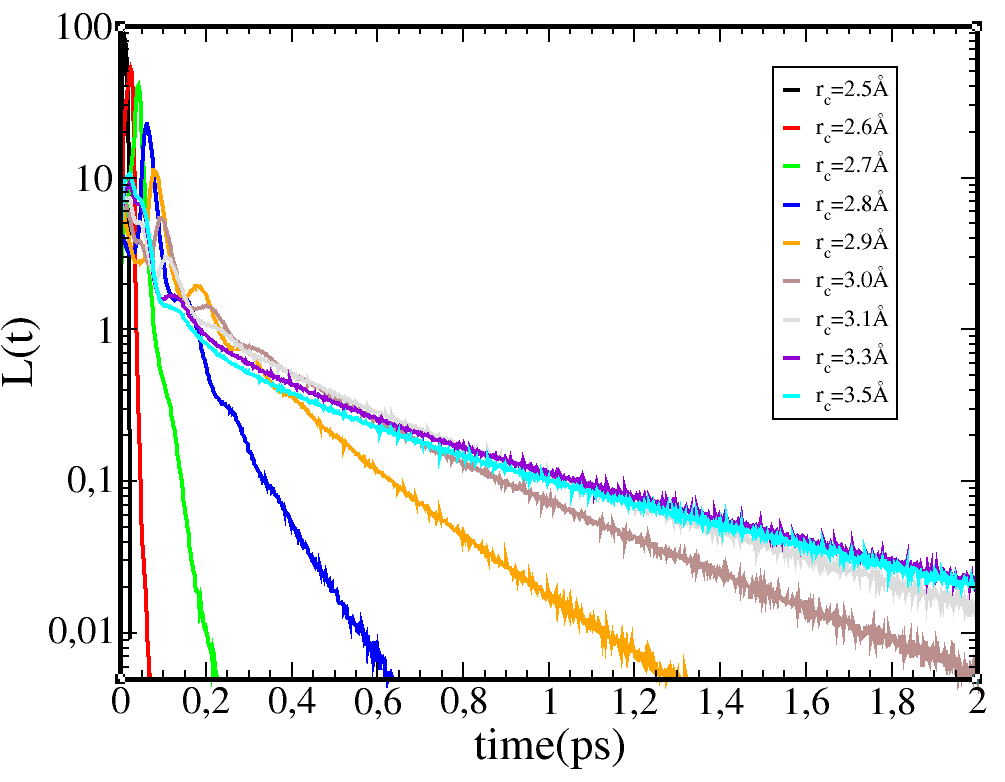} 
\caption{Long time behaviour of the H-bond life time distribution $L(t)$ for the
SPC/E model.}
\label{fig4-longtime}
\end{figure}

We now in position to confront our findings with respect to that of Luzar and Chandler
\cite{Luzar1996}. Fig.4 shows the long time behaviour of $L(t)$. It is readily
seen that, as the H-bonding distance is increased, the fast exponential
decay corresponding to smaller H-bonding distances converges towards
a slow, and possibly non-exponential behaviour, as witnessed by the
merging of the 3 curves corresponding to the range $r_{c}=3.1-3.5\mathring{A}$.
Luzar and Chandler have selected $r_{c}=3.5\mathring{A}$ in their
entire analysis, and this criterion seems to have been retained in
all subsequent literature \cite{SIM_Guardia_Wat_Alc,SIM_Skarmoutsos_Guardia_ethanol,SIM_EXP_TBAWat_Matija}.
In view of the present analysis, this approach would correspond to
a global, almost macroscopical view of the kinetics of association
in H-bonded liquids. While this may seem reasonable to make contact
with macroscopic physics, we show here that the microscopic physics
does not lead to a picture dominated by random distributions, but
on the contrary to a selective trinitary view of clustering.

What about other H-bonded liquids? These are usually not studied under
the same perspective as water. For example, H-bonding in water is often
discussed in terms of flickering clusters \cite{Frank1957} or network \cite{Perram1971}. These
specificities do not apply to other H-bonding liquids throughout the
literature. Is there a common clustering specificity to all H-bonding
liquids?

\subsubsection{Alcohols}

Fig.5 shows the H-bond lifetimes distributions $L(t)$ for the
OPLS methanol model. The comparative analysis for other models is
shown in the SI document. A comparison of Fig.5 with water in Fig.3
shows immediately that the same global features are present, namely
a high first peak which decreases to larger life times for larger
$r_{c}$ distances, while intermediate 2 life time peaks emerge at
distances around the attractive minimum of the mean force potential.

\begin{figure}[ht]
\centering \includegraphics[height=7cm]{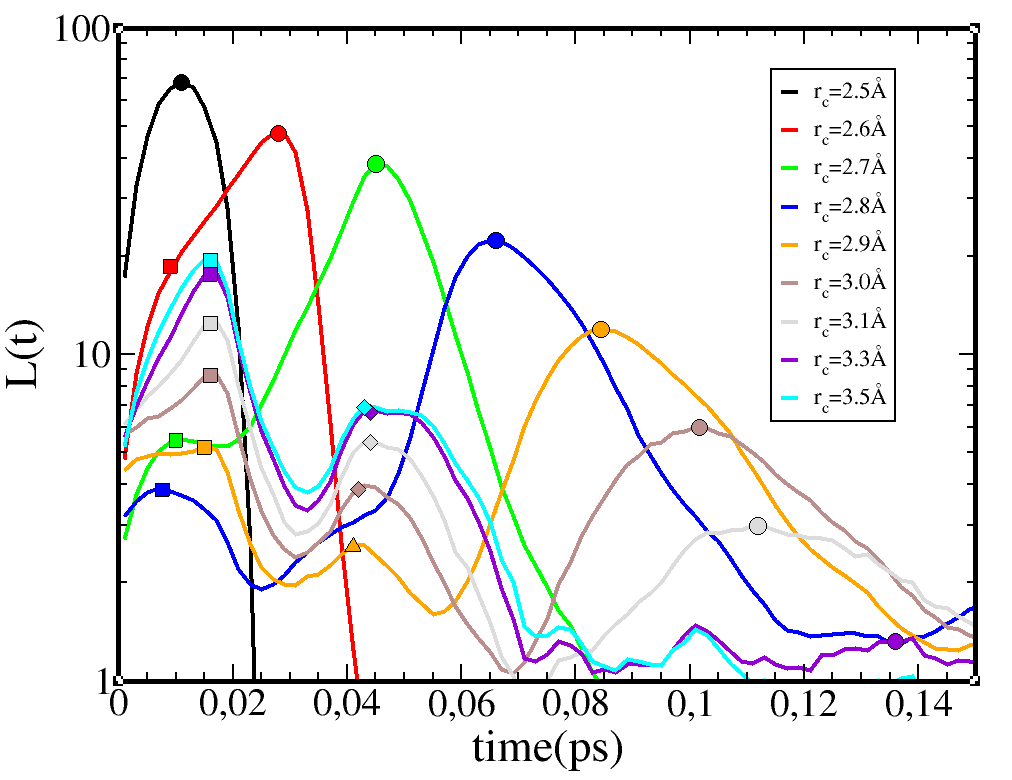} 
\caption{H-bond lifetime distribution for OPLS methanol, with different H-bonding
distance $r_{c}$. Symbols and line colors are same as in Fig.3 }
\label{fig5-meth}
\end{figure}

The principal differences are seen in the values of the maximum amplitudes
and positions, which are model dependent to a large extent as one
should expect for underlying differences in interactions, but also
in the secondary peaks. These latter peaks appear as more marked for
methanol than for water, and this is more particularly true for
the second peak, which is a true peak for the alcohol, while it was
more of a shoulder in the case of water. Another similarity with water
is the relative insensitivity of the positions of the second
and third peaks to the H-bonding $r_{c}$ distance. The fact that the
secondary peaks are more marked is in line with the known fact that
alcohols form linearly shaped clusters \cite{ExpScattMaginiMethanol,ExpScattNartenEthMeth,ExpScattFinnsMonools,
ExpScattMatijaMonools}
of different topology: chains, loops, lassos \cite{EXP_Sarkar_Joarder_Methanol,EXP_Sarkar_Joarder_ethanol,
ExpScattJoarderAlcohols,EXP_Pustai_MethEthProp,ExpScattBenmoreEthanol}.
These shapes are not found in water, at least in computer simulation
of popular water models. Therefore, the similarity of the peak cannot
refer to topological specificity, but to characteristic H-bonding patterns.
The cluster analysis of water models shows that water clusters are
either globular like or chain like. A similar analysis for the alcohols
indicate that linear and branched clusters are mostly present.

Fig.6a and Fig.6b show similar distributions for OPLS ethanol and
propanol, as well as Fig.6 of the SI for 1-octanol. 
Again, the similarity with both methanol and water is striking
and enforces the idea that the clustering features of the hydroxyl
group are pretty much the same across different H-bonding liquids bearing
this group, such as water and mono-ols.

\begin{figure}[ht]
\centering \includegraphics[height=4cm]{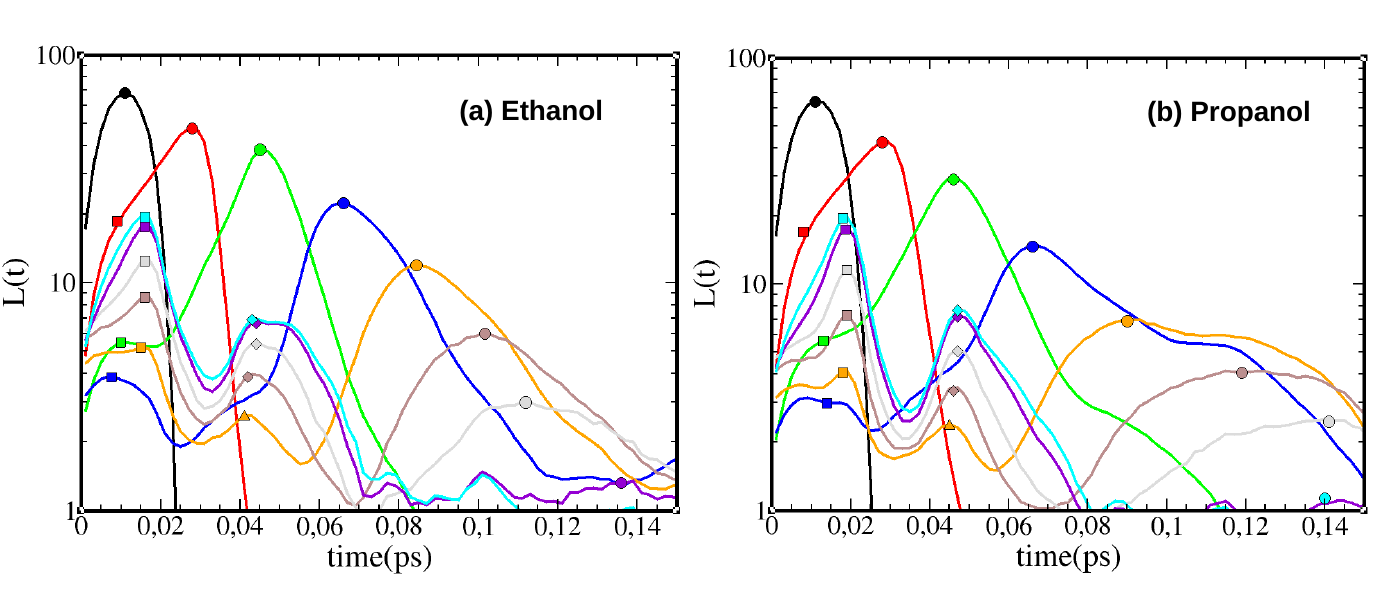} 
\caption{H-bond lifetime distribution for OPLS ethanol (a) and propanol (b),
with different H-bonding distance $r_{c}$. Symbols and line colors
are same as in Fig.3}
\label{fig-eth}
\end{figure}

\begin{figure}[ht]
\centering \includegraphics[height=7cm]{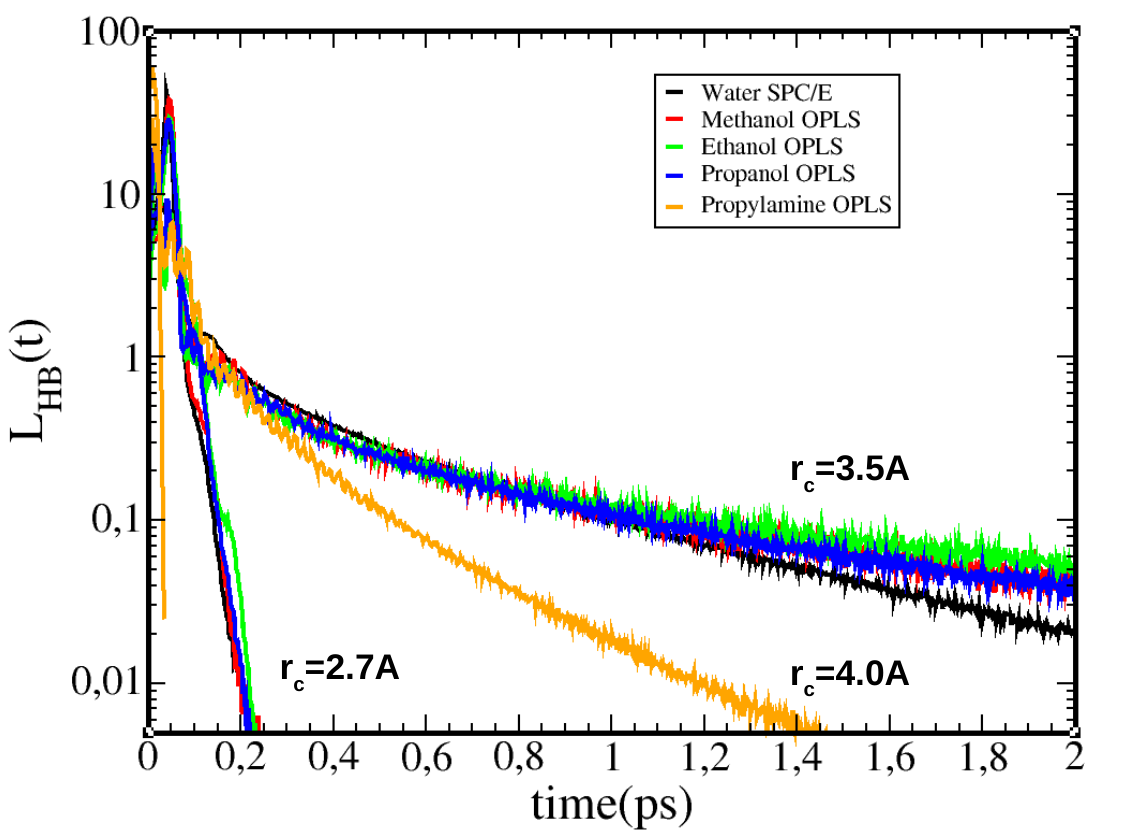} 
\caption{Long time behaviour of the H-bond lifetime distribution for all
the models studied herein}
\label{fig7-prop-2}
\end{figure}

The case of 1-propylamine is illustrated in Fig.7 of the SI, and again shows very similar
features for L(t), although this is a very different H-bonding liquid
with the amine group NH2, and an acceptor nitrogen atom which is
larger than the oxygen atom of water and hydroxyl groups of alcohols.
Hence, the distances associated to each curves are different than those in Figs(3,4).
This particular liquid confirms that the short time universality for L(t) uncovered
herein is a real feature of H-bonding liquids in general.

Similarly to Fig.4 for water, we compare in Fig.7 the long time decay of L(t)
for several H-bonding liquids, and
this for two different values of $r_c$. Interestingly, we find that for a given
$r_c$, all curves tend to lie quite close to one another, and that there is also
a species dependance. This is illustrated for $r_c=3\mathring{A}$ for water and
alcohols. We also note that, since 1-propylamine is a nitrogen based H-bonding liquid, 
its long time behaviour cannot be compared with that of oxygen atom based ones.
It is interesting to compare the long time analysis of $L(t)$ with the approach of
Luzar-Chandler which is based on the analysis of $c(t)$ and the related H-bond kinetics.
While the long time behaviour points to differences in Hbonding liquids, 
the short time transient part, which we study here demonstrates, that
the underlying transient dynamics are based on universal elementary
cluster structures. This is perhaps the main message of the present
findings.

\subsection{A test with the ``weak-water'' model}

In order to test the present conclusions, and in particular the inference methodology, 
we have studied previously introduced models of ``weak-water'' \cite{Kezic2013}. 
This model is based on the
SPC/E water for water, where the partial charges on the oxygen and
hydrogen atoms are scaled by a parameter $\lambda$ ($0<\lambda<1$),
allowing to tune the hydrogen bonding from the original model (with
$\lambda=1$) to a simple Lennard-Jonesium (with $\lambda=0$). It
was found that from $\lambda\le0.6$ the influence of partial charges
and hydrogen bonding were not relevant and the model was structurally
similar to a simple Lennard-Jones liquid. This model appears here
as a useful way to measure the cluster hypothesis for the complex
time dependence of $L(t$). As $\lambda$ is made smaller, the hydrogen
bonding abilities decrease, and it is possible to test directly the
influence of hydrogen bonding clusters on the shape of $L(t)$. 

In the order to preserve the liquid state for small $\lambda$ values
and under ambient conditions, it was found necessary to increase the
Lennard-Jones energy parameter $\epsilon=\epsilon(\lambda)$ according
to the decrease of $\lambda$. In the present test, we have bypassed
this procedure by doing the test simulations in the NVT Canonical
ensemble, hence keeping the volume fixed at that of the real liquid
water.

Since the structure of the weak-water liquid is strongly be affected
by the decrease of the partial charges, the H-bonding distances
must be adjusted appropriately. Fig.8a
shows the various $g_{OO}(r)$ for different $\lambda$ values we
have used here, namely $\lambda=0.8$, $0.5$ and $0.2$. The selected
bonding distances depend on the position of the maximum, and
differ quite a bit from that of the initial SPC/E water, ranging now
from $2.7\mathring{A}$ to $4.5\mathring{A}$. It is important to
note that, while we vary the bonding distance criteria, we keep the
angular criterion the same as that for pure water, which is that the
angle O-H-O is $180\pm30$. This means that, even though weaker water
model may have different bonding angles, we will still select only
a sub class of those obeying tetrahedral bonding directions. This
is justified by the fact that these models do not have as strong H-bonding
tendencies as SPC/E water, hence the angular bias is less significant
than the distance of bonding. But we will need to take into this bias
in order to interpret the data.

\begin{figure}[ht]
\centering \includegraphics[height=7cm]{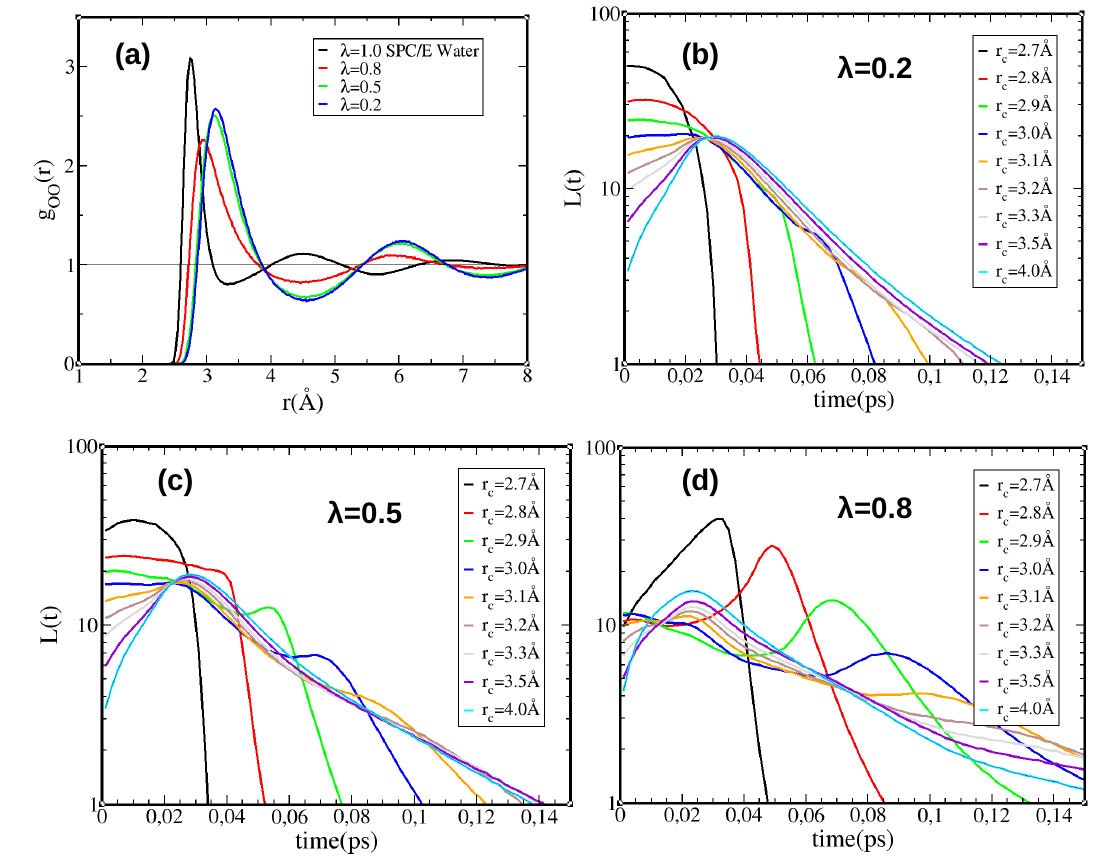} 
\caption{Weak water models data. (a) Comparison of the RDF of various models;
(b) H-bond lifetimes for $\lambda=0.2$; (c); (b) H-bond lifetimes for
$\lambda=0.5$; (d) H-bond lifetimes for $\lambda=0.8$. }
\label{fig8-1}
\end{figure}

Fig.8b shows the H-bond lifetimes for $\lambda=0.2$, which is very
close to a Lennard-Jones system. For very small bonding distances
$r_{c}\leq3.0\mathring{A}$ , we note that the $L(t)$ do not exhibit
a clear maximum, and sometimes even show a plateau-like behaviour,
such as for $R=3.0\mathring{A}$(blue curve). This is a direct consequence
of the angular bias described above. However, from bonding distance
$r_{c}\geq3.1\mathring{A}$, we observe a clear maximum, and this
maximum is nearly the same for all subsequent $r_{c}$ values. 
 This is what we
would expect in a quasi-LJ system, where pairing is nearly isotropic.
We also note that this lifetime distribution is not necessarily about
isolated pairs, and could involve those in larger clusters. We note
the presence of intriguing small shoulder-like features at large times,
such as for the green, blue and yellow curves, but these features
cannot be interpreted from Fig.8b alone, but will become clearer from
the analysis of the next cases.

Fig.8c shows the H-bond lifetimes for $\lambda=0.5$, which has the
same pair distribution as the $\lambda=0.2$ case, as seen in Fig.8a.
We note that most of the features observed previously equally appear
here - supporting the structural analogy, but the previously noted
intriguing shoulder structures has now grown into peak structures and
are very apparent. However, it still remain difficult to interpret
them by simple inference.

Fig.8d is for the case $\lambda=0.8$, which is the closest to SPC/E
water, and should be compared with Fig.3. 
The previous intriguing peaks have now grown to invade the entire
figure, and, by comparison with Fig.3, represent the dominant contribution
to the H-bond lifetimes for each R
$r_c$ value, as we have interpreted them when
discussing Fig.3. But now, we can finally understand the smaller features in Fig.8c,
 namely the origin of the secondary peaks, precisely because these were the dominant features
in the previous Figs.8b-c. Indeed, we have interpreted them as peaks related to bonded
pairs within clusters.
This is precisely the conclusion we have reached
when discussing Figs.3-6, but by inference. To summarize, the study
of the weak-water models allow us to confirm that the secondary peak
features are indeed related to clusters. 

\subsection{Influence of the alkyl tails}

Although the present study reveals similarities in the H-bonding life
times, one would expect that the presence of non-bonding alkyl tails
would affect the lifetime. This is illustrated in Fig.9 for all the
liquids studied in this work, water, methanol to 1-octanol and 1-propylamine.
Since the alkyl tail influence is best seen at largest distance cutoff
we have considered $r_c=3.5\mathring{A}$ for all systems except for
propylamine, for which we have used appropriately $r_c=4.0\mathring{A}$.

\begin{figure}[ht]
\centering \includegraphics[height=7cm]{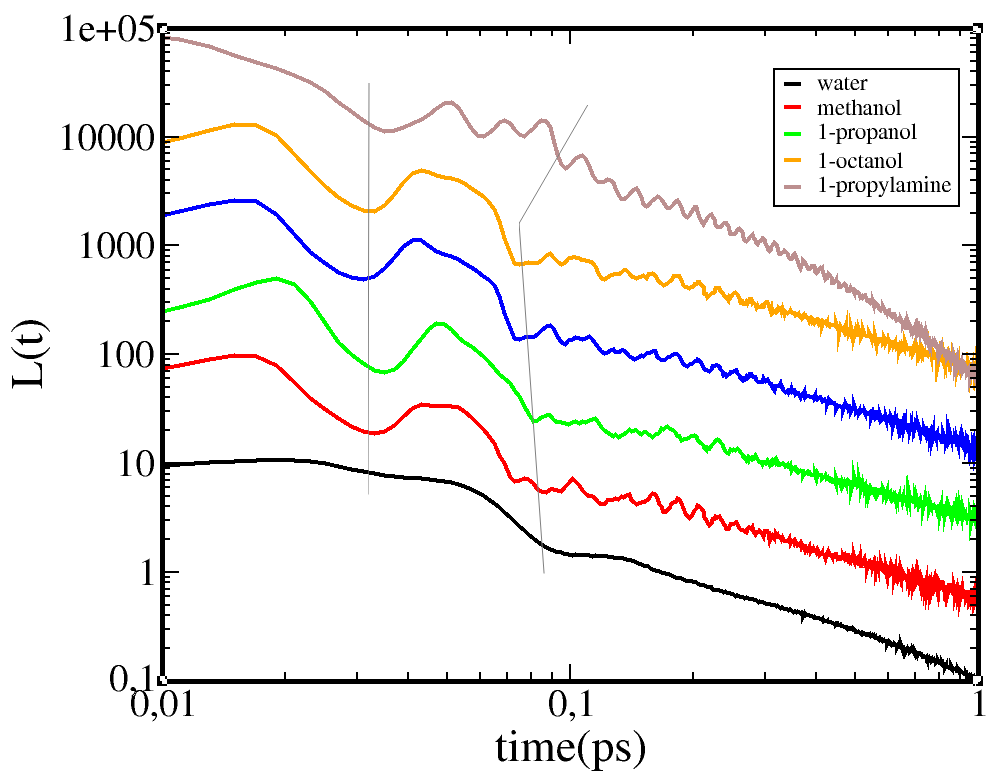} 
\caption{
Influence of alkyl tails on lifetime distribution for methanol (red), ethanol (green), propanol(blue), octanol (gold) and propylamine (purple). The black curve is for water and serves as a reference. The grey lines serve to delimitate the 2nd and 3rd peaks regions mentioned in the text.
}
\label{fig9}
\end{figure}
For clarity, each curve has been shifted by log(10) from the previous
one. The bottom black curve is for water, and serves only as a reference, since
water has no alkyl tail, and consequently does not show
anything particular other than the features discussed above. However, all
the alcohol curves show tiny oscillations past the 3rd peak. Higher alcohols
such as 1-propanol and 1-octanol have even their 3rd peak weakly modulated.
We attribute these oscillations in lifetime to the presence of the alkyl
tail "bath", which surrounds the OH clusters, and affects the decay of their
life time. This interesting feature further supports the interpretation that
the 3rd peak is associated with the evolution of the cluster topology in time.
This feature is naturally absent from water because there are no alkyl
groups. However, the case of the propylamine (the upper most curve in 
purple) is very interesting. In previous studies \cite{2017_Propylamine1}, we have compared the 
clustering of neat 1propanol to that of 1-propylamine, and shown that the
clustering of the amine group was not so important, both in size and shape,
as the chain patterns observed for 1propanol. The alkyl chain, being the
same between the 2 species, plays an indirect role in this incomplete
clustering of the NH2 amine groups. What we observe in Fig.9 is that the
role of the alkyl tail is so important that it "bites" into the 3rd peak
which is about cluster topology, and makes even this peak look less
characteristic compared to those of the alcohols. These oscillations also
last longer than in the case of alcohols.
This figure reveals the dynamical role played by the seemingly neutral alkyl
tail background. The role of these tails was emphasized in a previous study
\cite{2020_Neat_Alc_Germans} of the shape of the Xray pre-peak feature of alcohols. This role is
confirmed through the present study.

\section{Discussion and Conclusion}

While chemical matter, and more importantly bio-chemical matter, is 
known to be made of labile objects, in addition to well defined atoms and
molecules. It raises the question whether
such fleeting objects could play a role as important as the permanent
ones. The pertinence of this question is enforced by the recognition
that liquid assemblies of atoms and molecules are subject to fluctuations
which depend on the nature of the interactions, and in particular
highly directional ones such as the H-bond interaction. Self assembly
is the driving mechanism of many complex systems, such as in soft-matter
and biological systems. The microscopic interaction at the root of
self-assembly is the H-bonding process. This mechanism can be relatively
well captured by classical force field simulation using Coulomb charge
pairing. 

The calculations reported herein show that, for a given H-bonding distance
$r_c$, the H-bond lifetime is essentially dominated by one mean lifetime
for distances $r_c$ smaller than the first peak of $g(r)$, but that
two secondary peaks appear for larger distances until the first minimum
of $g(r)$, and that these become dominant as $r_c$ is increased.
We have attributed these peaks to linear and non-linear cluster geometries.
In addition, the present study reveals a previously unexpected similarity
in H-bonding lifetime distribution in the small distance and short
time regime corresponding to distances under $3.5\mathring{A}$ and
under 150 fs, and this independently of the nature of the H-bonding group.
We have provided convincing
empirical arguments for the existence of three type of H-bond based
clustering specificities in three different associated liquids, water,
mono-ols and amines. These arguments suggest that there exist two
distinct families of clusters, inside which the dimer lifetime plays
a very different role. The first family concerns linearly shaped clusters,
whether these are line-like or circular. The second family, we have
refered as non-linear clusters, which covers all other forms, such
as globular -as those found in water, or branched, such as lasso-shaped
found in mono-ols \cite{Guo2003,2020_Neat_Alc_Germans}. 

The calculations concern primarily lifetime of pair of H-bonded particles.
But such pairs often reside inside specific clusters, hence their
lifetime is affected by the life of the entire cluster. Because of
the cooperative motions of particles linked within a cluster, we expect
that the these motions affect the lifetime statistics in particular
ways, as to emerge the 3 peaks observed in the data reported herein. 

\section*{Acknowledgments}

This work has been supported in part by the Croatian Science Foundation
under the project UIP-2017-05-1863 ``Dynamics in micro-segregated
systems''.

\bibliographystyle{plain}

\end{document}